\documentclass[structabstract,rnote]{aa}  
\usepackage{amssymb}
\usepackage{graphicx}
\usepackage{epsfig}
\usepackage{subfigure}
\usepackage[authoryear]{natbib}
\usepackage[figuresright]{rotating}

\usepackage{txfonts}
\newcommand{\teff}{\ifmmode T_{\rm eff} \else T$_{\mathrm{eff}}$\fi}
\newcommand{\logg}{\ifmmode \log g \else $\log g$\fi}
\newcommand{\lL}{\ifmmode \log \frac{L}{L_{\odot}} \else $\log \frac{L}{L_{\odot}}$\fi}

\newcommand{\kms}{km s$^{-1}$}
\newcommand{\msun}{\ifmmode M_{\odot} \else M$_{\odot}$\fi}
\newcommand{\zsun}{\ifmmode Z_{\odot} \else Z$_{\odot}$\fi}
\newcommand{\lsun}{\ifmmode L_{\odot} \else L$_{\odot}$\fi}
\newcommand{\rsun}{\ifmmode R_{\odot} \else R$_{\odot}$\fi}
\newcommand{\qh}{\ifmmode Q_{\rm H} \else $Q_{\rm H}$\fi}
\newcommand{\qhei}{\ifmmode Q_{\ion{He}{i}} \else $Q_{\ion{He}{i}}$\fi}
\newcommand{\mum}{$\mu$m}

\newcommand{\brg}{Br$\gamma$}

\begin{document}
   \title{Near-Infrared spectroscopy of the super star cluster in NGC1705 \thanks{Based on observations collected at the ESO/VLT under program 384.D-0301(A).}}

   \subtitle{}

   \author{F. Martins\inst{1}
          \and
          N.M. F$\rm{\ddot{o}}$rster Schreiber\inst{2}
          \and
          F. Eisenhauer\inst{2}
          \and
          D. Lutz\inst{2}
          }

   \offprints{F. Martins}

   \institute{LUPM--UMR5299, CNRS \& Universit\'e Montpellier II, Place Eug\`ene Bataillon, F-34095, Montpellier Cedex 05, France\\
              \email{fabrice.martins AT univ-montp2.fr}
         \and
             Max Planck Institute for extraterrestrial Physics, PO Box 1312, Giessenbachstr., 85741 Garching, Germany \\
             }

   \date{Received ...; accepted ...}

 
  \abstract
   {}
   {We study the near--infrared properties of the super star cluster NGC1750--1 in order to constrain its spatial extent , its stellar population and its age.}
   {We use adaptive optics assisted integral field spectroscopy with SINFONI on the VLT. We estimate the spatial extent of the cluster and extract its K--band spectrum from which we constrain the age of the dominant stellar population.}
   {Our observations have an angular resolution of about 0.11\arcsec, providing an upper limit on the cluster radius of 2.85$\pm$0.50 pc depending on the assumed distance. The K--band spectrum is dominated by strong CO absorption bandheads typical of red supergiants. Its spectral type is equivalent to a K4--5I star. Using evolutionary tracks from the Geneva and Utrecht groups, we determine an age of 12$\pm$6 Myr. The large uncertainty is rooted in the large difference between the Geneva and Utrecht tracks in the red supergiants regime. The absence of ionized gas lines in the K--band spectrum is consistent with the absence of O and/or Wolf--Rayet stars in the cluster, as expected for the estimated age.  }
    {}

   \keywords{Galaxies: clusters: individual: NGC1705--1 -- Stars: massive -- Stars: late-type}

   \maketitle


\section{Introduction}
\label{s_intro}

Super star cluster are found in various galaxies: starburst galaxies \citep[M82,][]{oc95,mccrady03,gs99}, interacting galaxies \citep[NGC4038/39, the Antennae,][]{whitmore95}, amorphous galaxies \citep[NGC1705][]{oc94}. Compared to the most massive clusters found in the Galaxy and the Magellanic Clouds, they have larger estimated masses \citep[in excess of 10$^{5}$ \msun\ and usually closer to a few 10$^{6}$ \msun,][]{mengel02,larsen04,bastian06}. Their mass distribution follows a power law with an index equal to $-$2 \citep{fall04}. \citet{fz01} have shown that a distribution of young massive clusters with such a mass function could evolve into a lognormal mass function similar to that of old globular clusters. Super star clusters may thus be the progenitors of globular clusters, although this is a debated question. Among the difficulties of this scenario, super star clusters have to survive the so--called ``infant mortality''. \citet{fall05} showed that the distribution of the number of clusters as a function of time in the Antennae galaxies was dramatically decreasing: $dN/d\tau \propto\ \tau^{-1}$ where $N$ is the number of clusters and $\tau$ the age. Either clusters are born unbound and dissolve rapidly, or they experience negative feedback effects from the most massive stars. Supernovae explosions and stellar winds can expel interstellar gas on short timescales leading to the cluster disruption \citep{gb06}. The way this feedback affects the cluster's evolution depends on the stellar content and its distribution. If the stellar initial mass function is top--heavy \citep[as seems to be the case in some clusters,][]{sternberg98} the presence of a large number of massive stars will enhance disruption. But if the massive stars are concentrated in the cluster core due to initial or dynamical mass segregation, their effects might be reduced. Information of the stellar content is thus necessary to better understand the evolution of these mini--starbursts. 

In this paper, we present new observations of the super star cluster in the amorphous galaxy NGC1705 \citep{melnick85a}. NGC1705--1 is one of the brightest clusters \citep[M$_{V}$=-15.4][]{ma01}. It is also one of the closest, at a distance of 5.1$\pm$0.6 Mpc \citep{tosi01}. Measuring velocity dispersions and assuming a bound cluster, \citet{hp96} determined a mass of 8.2$\pm$2.1 10$^{4}$ \msun. Using a larger gravitational radius, \citet{sternberg98} obtained M=2.7 10$^{5}$ \msun. \citet{sternberg98} used the L/M ratio derived from photometry and velocity dispersion to constrain the initial mass function. He found that the IMF is either flatter than the Salpeter IMF (slope $<$ 2.0) or that it is truncated at masses below 1--3 \msun. \citet{sg01} confirmed the latter conclusion and \citet{vazquez04} showed that a lower mass limit of 1 \msun\ for a Salpeter IMF was still compatible with the observed luminosity to mass ratio. \citet{melnick85a} showed that the optical spectrum of NGC1705--1 was typical of early B stars, excluding the presence of hotter objects such as O and/or Wolf--Rayet stars. This places a lower limit to the age of the cluster to 8--10 Myr. In a subsequent study, they detected the presence of CO bandheads in near--infrared narrow band photometry. Such features are typical of evolved cool stars such as red supergiants or AGB/RGB stars \citep{melnick85b}. \citet{hp96} obtained high resolution optical spectra of NGC1705--1 and confirmed the presence of both early B stars signatures (from spectral lines below 4500 \AA) as well as red supergiants metallic lines \citep[above 4500 \AA, see also][]{meurer92}. The presence of early B stars was confirmed by the UV spectra of \citet{vazquez04} which are typical of B0--1 V/III stars. Such an average spectral type corresponds to a dominant population of hot stars of age 12$^{+3}_{-1}$ Myr. \citet{marlowe95} estimated an age of 10 to 16 Myr for the starbust event in NGC1705 using UBV colors and the H$\alpha$ flux in comparison with starburst models. The metallicity of the host galaxy NGC1705 is subsolar with Z=0.35Z$_{\odot}$ \citep{ls04}. 

In this research note, we present the first near--infrared spectrum of the super star cluster NGC1705--1 obtained from integral field spectroscopy with SINFONI on the ESO/VLT. We provide an upper limit on the spatial extent of the cluster and determine its K--band spectral type. We give age estimates and discuss their uncertainties.


\section{Observations and data reduction}
\label{s_obs}

The observations were performed with the integral field near--infrared spectrograph SINFONI \citep{spiffi,bonnet04} on the ESO/VLT in service mode between December 6$^{th}$ 2009 and January 22$^{nd}$ 2010. The seeing was usually good, between 0.7 and 1.0\arcsec. We used the adaptive optics system with the cluster itself as guide star. We used both the 100mas and the 25mas plate scale in order to probe the very cluster as well as its immediate surrounding. Table \ref{tab_obs} provides the journal of observations. 

Data reduction was performed with the SPRED software \citep{spred}. After flat field correction and bias/sky subtraction, wavelength calibration was done using Ne--Ar lamp calibration data. Fine tuning based on atmospheric features provided the final wavelength calibration. Telluric lines were removed using early B stars spectra taken just after the science data and from which the stellar Br$\gamma$ (and \ion{He}{i} 2.11 \mum\ feature when present) were corrected. 

\begin{figure*}[]
\centering
\includegraphics[width=14cm]{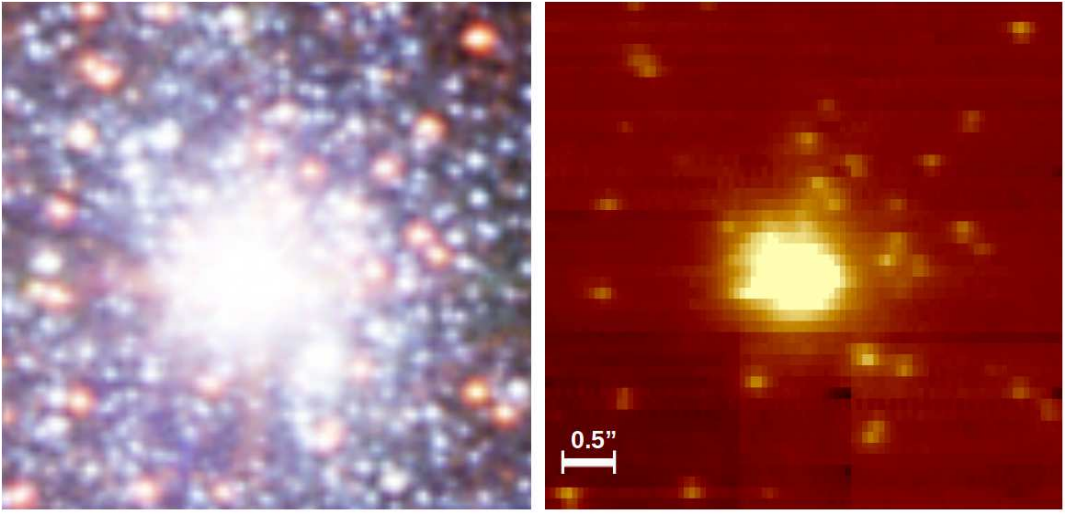}
\caption{{\it Left:} HST UBVI composite image of the super star cluster and its close environment (Tosi et al., Hubble Heritage Team -- STScI/AURA, NASA, ESA). {\it Right:} combination of all three SINFONI mosaics obtained with the 100 mas pixel scale frame. } \label{hst_vlt}
\end{figure*}

To estimate the spatial resolution of our data, we observed two point source (stars) of similar magnitude to NGC1705--1 immediately after the observations of the cluster on Jan. 10$^{th}$ and Jan. 11$^{th}$ 2010. These data are used to derive the width of the PSF. Fig.\ \ref{hst_vlt} shows our 100 mas pixel scale mosaic (right) together  with an HST UBVI composite image at the same scale (left).

\begin{table}
\begin{center}
\caption{Journal of observations. } \label{tab_obs}
\begin{tabular}{lcccc}
\hline
Date  & pixel scale & exposure time & seeing  & airmass\\
      & [mas]       & [s]           & \arcsec &        \\
\hline
& & NGC1705--1 & & \\
06 dec 2009      & 25  & 4$\times$100  &   0.81--0.94 & 1.22--1.26 \\
23 dec 2009      & 25  & 4$\times$100  &   0.60--0.75 & 1.40--1.55 \\
10 jan 2010      & 100 & 4$\times$100  &   0.56--1.00 & 1.17--1.22 \\
11 jan 2010      & 25  & 4$\times$100  &   0.96--2.60 & 1.23--1.31 \\
22 jan 2010      & 100 & 4$\times$100  &   0.83--1.13 & 1.18--1.20 \\
\hline
& & PSF calibrator & & \\
10 jan 2010      & 100  & 2$\times$60  &   0.68--0.75 & 1.24--1.24 \\
11 jan 2010      & 25  & 2$\times$60  &   0.95--0.99 & 1.38--1.38 \\
\hline
\end{tabular}
\end{center}
\end{table}

\section{Spatial distribution}
\label{s_spatial}

We have used the 25mas pixel scale mosaics taken on December 9$^{th}$ 2009, December 23$^{rd}$ 2009 and January 11$^{th}$ 2010 to estimate the spatial extent of NGC1705--1. Performing 2D Gaussian fits to the data, we obtain the following values for the Full Width at Half Maximum:  0.112\arcsec$\times$0.114\arcsec, 0.110\arcsec$\times$0.121\arcsec, 0.132\arcsec$\times$0.150\arcsec. On January 11$^{th}$ 2010, we observed a standard star used as a PSF calibrator. The 2D Gaussian fit gives a 2D FWHM of 0.102\arcsec$\times$0.111\arcsec. These measurements indicate that the core of NGC1705--1 is not resolved by our observations. The variations of the cluster FWHM from night to night are mainly due to varying seeing conditions. On January 11$^{th}$ 2010, the PSF is about 30\% smaller on the standard star, but the average seeing was also smaller during the observation compared to the cluster observation (see Table \ref{tab_obs}).

\begin{figure}[]
\centering
\includegraphics[width=9cm]{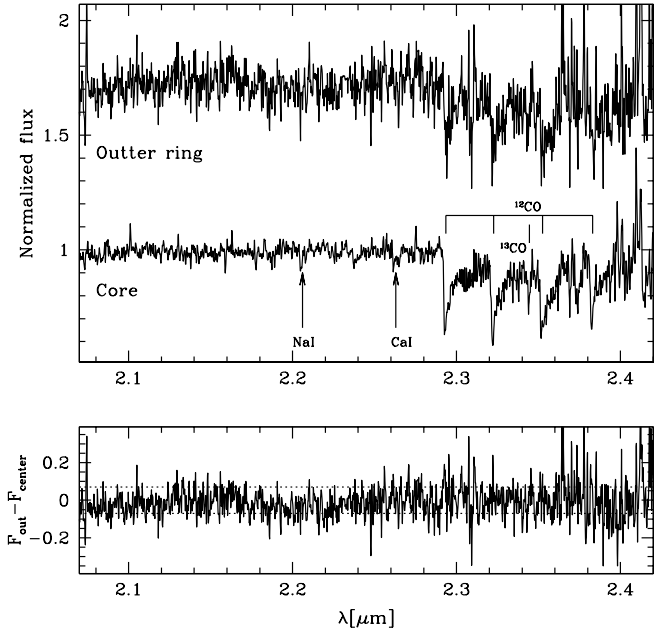}
\caption{Comparison between the spectrum extracted from a region of radius $\sim$0.12\arcsec\ centered on the cluster core and the spectrum extracted from a ring of radius 0.13--0.18\arcsec. The bottom panel shows the difference between both spectra. The dotted lines indicate the 1$\sigma$ deviation.} \label{fig_core_out}
\end{figure}

In Fig\ \ref{fig_core_out}, we show the spectrum obtained in two different regions of the 25mas mosaic: a circle centered on the cluster core and of spatial radius $\sim$ 0.12\arcsec, and a ring--like region located between $\sim$0.13\arcsec and $\sim$0.18\arcsec. The resulting spectra are shown in the bottom panel. The main lines are indicated. There is very little difference between the two spectra (see the plot of the difference as a function of wavelength in the bottom panel). This most likely indicates that we are observing the far wing of the PSF in the annulus region, confirming that we are not resolving the cluster with our observations. We can provide upper limits on its half--light radius. According to the values of FWHM given above, the cluster core is smaller than 0.11--0.12\arcsec. At the distance of NGC1705, this corresponds to a physical radius of less than 2.85$\pm$0.50 pc \citep[using the dispersion of the FWHM measurements as error on the angular size and for the distance of 5.1$\pm$0.6 Mpc of][]{tosi01}. 

\citet{oc94} determined a half--light radius of 0.14\arcsec\, corresponding to 3.4 pc using a distance of 5.0$\pm$2.0 Mpc. \citet{meurer95} found a significantly smaller value (0.04\arcsec, 1.1pc) using the same set of HST/WFPC data. \citet{sg01} concluded from their HST/WFPC2 observations that the half--light radius was 1.6$\pm$0.4 pc for a slightly larger distance (5.3$\pm$0.8 pc). Our \textit{upper limit} on the cluster size is consistent with the largest values derived previously, and compatible with the small radius quoted by \citet{meurer95} and \citet{sg01}. The size of the cluster NGC1705--1 remains poorly constrained at present, and future observations with ELTs and/or JWST are necessary to probe the spatial structure of this (and other) super star cluster.

\section{Cluster age}
\label{s_age}

\begin{figure}[]
\centering
\includegraphics[width=8cm]{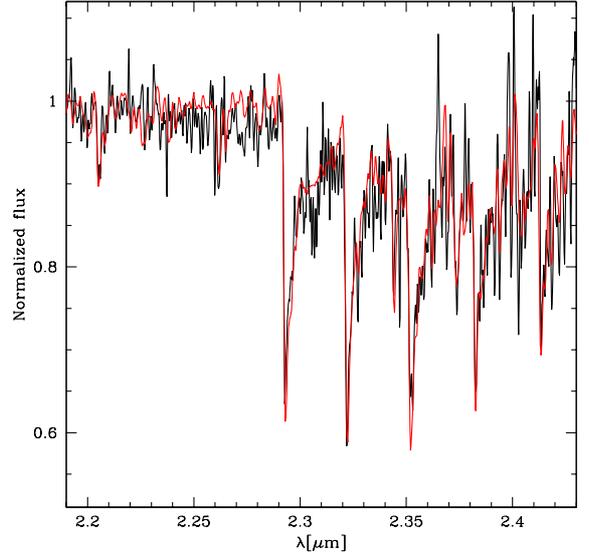}
\caption{Comparison between the K band spectrum of NGC1705--1 (black) and the spectrum of the Galactic K4.5Ib red supergiant HD~78647 (red). The latter spectrum is taken from the atlas of \citet{wh97}. The SINFONI spectrum has been degraded to the resolution of the template spectrum (R=2000).} \label{comp_spec}
\end{figure}

We first note the absence of Br$\gamma$ emission in the spectrum of NGC1705--1 (we determine an upper limit on the \brg\ equivalent width of 0.1 \AA), consistent with the absence of hot massive stars and thus of a very young population. As seen above, CO bandhead absorption dominate the K--band spectrum, as in late type stars. We have compared by eye the cluster spectrum with templates spectra of cool, evolved stars taken from the atlas of \citet{wh97}. We find that the former is best accounted for by the spectrum of a K4.5Ib star (Fig.\ \ref{comp_spec}). For later spectral types, the CO bandheads are too deep. A supergiant luminosity class is also preferred, since giant stars spectra do not have broad enough CO overtones \footnote{Note that the Wallace \& Hinkle atlas does not contain all spectral types and luminosity classes. Hence, the best representative spectral type should not be trusted at the level of a sub-spectral type.}. This result is confirmed by the calculation of the first CO overtone equivalent width (38.2 \AA\ measured between 2.2900 and 2.3200 \mum). From Fig.\ 8 of \citet{rsg1}, in which the CO equivalent width was measured on the same interval as us, we see that such an equivalent width is observed in K3--4I stars, as well as marginally in M5--6 giants. The two independent estimates favor a K4--5I spectral type for the entire cluster which is thus dominated by the near--infrared light of red supergiants. If we assume that the supergiants of NGC1705--1 are in their coolest evolutionary state, the fact that their spectral type is K and not M is an indication that the metallicity of the cluster is sub-solar. As shown by \citet{mo03}, the distribution of spectral types among red supergiants shifts towards earlier spectral types when metallicity decreases. While M2I is the spectral type the most represented in the Galaxy, M1I stars populate in majority the LMC and K5--7I the SMC. This is consistent with the study of \citet{meurer92} and \citet{storchi94} who report a sub--solar global metallicity for the entire galaxy NGC~1705: the former derive 12+log(O/H)=8.46, while the second give 8.36. These values are similar to the LMC (respectively SMC) metallicity. 

\begin{figure}[]
\centering
\includegraphics[width=8cm]{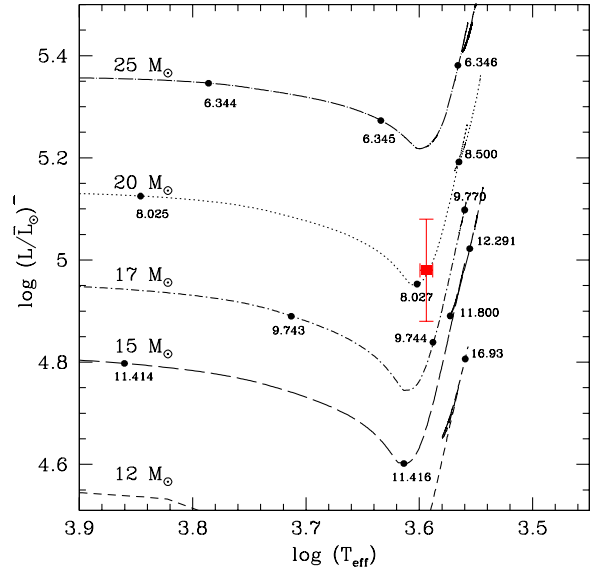}
\caption{HR diagram showing the position of an SMC K4--5I star. The evolutionary tracks are from \citet{brott11a} and have an initial rotational velocity of 300 \kms\ and the SMC composition. The dots on the tracks indicate the ages in Myr. The uncertainty on the red supergiant parameters reflect the range of values determined for SMC K4--5I stars in the study of \citet{levesque06}. } \label{hr_ssc}
\end{figure}

This estimate of the dominant population in NGC1705--1 can be used to derive the age of the cluster. \citet{vazquez04} proceeded this way to report an age of 12$^{+3}_{-1}$ Myr using a HST/STIS UV spectrum of the cluster. In Fig.\ \ref{hr_ssc} we show the position of a typical SMC K4--5 supergiant in the HR diagram, using parameters from \citet{levesque06}, i.e.: \teff\ = 3925$\pm$50 K and \lL\=4.98$\pm$0.10. The evolutionary tracks of \citet{brott11a} are overplotted. They include rotational mixing (for an initial rotation rate of 300 \kms) and have the SMC composition. Ages are indicated in Myr by filled circles along the tracks. From that figure, we see that an age of about 7 to 10 Myr can be inferred for NGC~1705--1. If we were to use the LMC tracks and average LMC properties of K4--5I stars \citep[still from][]{levesque06} we would derive an age of 5.5 to 7.5 Myr. These numbers are significantly lower than the value reported by \citet{vazquez04}. The reason is the use of different sets of evolutionary tracks. \citet{vazquez04} relied on the non--rotating Geneva tracks, while we use the rotating Utrecht tracks. Using the non--rotating Geneva tracks, we find ages of 8--14 Myr and 12--17 Myr for the LMC and SMC cases respectively. These values are in much better agreement with that of \citet{vazquez04}. We can thus conclude that our results are consistent with theirs. But the main conclusion is that the choice of evolutionary tracks is crucial to establish the age of the cluster. Depending on the tracks used, systematic differences of the order of 50\% of the cluster age can be made. 

To further quantify the effects of evolutionary tracks on age determinations, we can compare the results obtained from the Geneva and Utrecht \textit{rotating} tracks.  At solar metallicity \footnote{Comparisons are not possible at LMC/SMC metallicities because Geneva rotating tracks are not available at those metallicities for stars with M$<$25 \msun.}, important differences in the behaviour of the tracks are found at \teff\ lower than 10000\,K. The Geneva tracks have an almost constant luminosity until the lowest temperatures where a luminosity increase happens. The Utrecht tracks show a decreasing luminosity until approximately 4000\,K before a rise at lower \teff. As a consequence, a star with \teff\ = 4000\,K and \lL\ = 4.80 is reproduced by the 20 \msun\ Utrecht track at 8.7 Myr, and by the 15 \msun\ Geneva track at 13.9 Myr. There is a 5 Myr difference in the age estimate. A detailed understanding of these differences is beyond the scope of this paper. It may be due to a different treatment of convection. The conclusion one can draw is that, as illustrated by our analysis of NGC1705--1, the ages derived using the Utrecht tracks are much lower than those determined with the Geneva tracks. Hence, an accurate age determination cannot be performed, not because of the quality of the observational data, but due to the uncertainties in the theoretical tracks.

We can use the equivalent width of the first CO overtone to get an independent estimate the age of the population \citep{mengel01}. Population synthesis models predict the evolution of the strength of this feature as a function of time, depending on the assumed star formation history, initial mass function, stellar libraries and isochrones. We measure an equivalent width of 12.8 \AA\ for the first CO overtone \citep[measured between 2.2924 and 2.2977 \mum\ according to][]{ori93}. Using the starburst models of \citet{leitherer99} (including non--rotating tracks, a Salpeter IMF, a burst of star formation) we see from their Fig. 101c that at a slightly sub--solar metallicity, the first CO overtone equivalent width (computed on the same interval as us) is in the range 10--16 \AA\ for ages between 7 and 30 Myr. This is consistent with our estimates.

In conclusion, performing an age determination for NGC1705--1 is found to be a difficult task given the current uncertainties on evolutionary models in the red supergiant phase. Based on our estimates, we can quote a value of 12$\pm$6 Myr. This is still compatible with the absence of ionized gas emission (\brg) that would be produced by a large population of ionizing sources (O and Wolf--Rayet stars).  

\section{Conclusion}
\label{s_conc}

We have presented the first near--infrared integral field spectroscopy of the super star cluster NGC1705--1 obtained with SINFONI on the ESO/VLT. The cluster is found to have an angular size smaller than about 0.11--0.12\arcsec\ and is not resolved by our AO--assisted observations. This places an upper limit of 2.85$\pm$0.50 pc (depending on the distance) on its radial extension. The K--band spectrum of the cluster is dominated by strong CO absorption bandheads. It is similar to the spectrum of a red supergiant of spectral type K4--5. There is no sign of ionized gas in the spectrum. This confirms previous studies indicating that the cluster contains massive stars, but no O and/or Wolf--Rayet objects. Using different evolutionary tracks, we estimate the age to be 12$\pm$6 Myr. The large uncertainty is rooted on the important differences between the Geneva and Utrecht evolutionary tracks in the supergiant regime, and not in the quality of the observational data. Depending on the type of tracks used, ages can systematically differ by 5--7 Myr.

\begin{acknowledgements}
We acknowledge the suggestions of an anonymous referee. We thank the ESO/Paranal staff for performing the observations in service mode. FM acknowledges support from the ``Agence Nationale de la Recherche''.
\end{acknowledgements}

\bibliography{biblio.bib}

\begin{thebibliography}{36}
\expandafter\ifx\csname natexlab\endcsname\relax\def\natexlab#1{#1}\fi

\bibitem[{{Abuter} {et~al.}(2006){Abuter}, {Schreiber}, {Eisenhauer}, {Ott},
  {Horrobin}, \& {Gillesen}}]{spred}
{Abuter}, R., {Schreiber}, J., {Eisenhauer}, F., {et~al.} 2006, New Astronomy
  Review, 50, 398

\bibitem[{{Bastian} {et~al.}(2006){Bastian}, {Saglia}, {Goudfrooij},
  {Kissler-Patig}, {Maraston}, {Schweizer}, \& {Zoccali}}]{bastian06}
{Bastian}, N., {Saglia}, R.~P., {Goudfrooij}, P., {et~al.} 2006, \aap, 448, 881

\bibitem[{{Bonnet} {et~al.}(2004){Bonnet}, {Abuter}, {Baker}, {Bornemann},
  {Brown}, {Castillo}, {Conzelmann}, {Damster}, {Davies}, {Delabre},
  {Donaldson}, {Dumas}, {Eisenhauer}, {Elswijk}, {Fedrigo}, {Finger},
  {Gemperlein}, {Genzel}, {Gilbert}, {Gillet}, {Goldbrunner}, {Horrobin}, {Ter
  Horst}, {Huber}, {Hubin}, {Iserlohe}, {Kaufer}, {Kissler-Patig}, {Kragt},
  {Kroes}, {Lehnert}, {Lieb}, {Liske}, {Lizon}, {Lutz}, {Modigliani}, {Monnet},
  {Nesvadba}, {Patig}, {Pragt}, {Reunanen}, {R{\"o}hrle}, {Rossi}, {Schmutzer},
  {Schoenmaker}, {Schreiber}, {Stroebele}, {Szeifert}, {Tacconi}, {Tecza},
  {Thatte}, {Tordo}, {van der Werf}, \& {Weisz}}]{bonnet04}
{Bonnet}, H., {Abuter}, R., {Baker}, A., {et~al.} 2004, The Messenger, 117, 17

\bibitem[{{Brott} {et~al.}(2011){Brott}, {de Mink}, {Cantiello}, {Langer}, {de
  Koter}, {Evans}, {Hunter}, {Trundle}, \& {Vink}}]{brott11a}
{Brott}, I., {de Mink}, S.~E., {Cantiello}, M., {et~al.} 2011, \aap, 530, A115+

\bibitem[{{Eisenhauer} {et~al.}(2003){Eisenhauer}, {Tecza}, {Thatte}, {Genzel},
  {Abuter}, {Iserlohe}, {Schreiber}, {Huber}, {Roehrle}, {Horrobin},
  {Schegerer}, {Baker}, {Bender}, {Davies}, {Lehnert}, {Lutz}, {Nesvadba},
  {Ott}, {Seitz}, {Schoedel}, {Tacconi}, {Bonnet}, {Castillo}, {Conzelmann},
  {Donaldson}, {Finger}, {Gillet}, {Hubin}, {Kissler-Patig}, {Lizon}, {Monnet},
  \& {Stroebele}}]{spiffi}
{Eisenhauer}, F., {Tecza}, M., {Thatte}, N., {et~al.} 2003, The Messenger, 113,
  17

\bibitem[{{Fall}(2004)}]{fall04}
{Fall}, S.~M. 2004, in Astronomical Society of the Pacific Conference Series,
  Vol. 322, The Formation and Evolution of Massive Young Star Clusters, ed.
  H.~J.~G.~L.~M. {Lamers}, L.~J. {Smith}, \& A.~{Nota}, 399

\bibitem[{{Fall} {et~al.}(2005){Fall}, {Chandar}, \& {Whitmore}}]{fall05}
{Fall}, S.~M., {Chandar}, R., \& {Whitmore}, B.~C. 2005, \apjl, 631, L133

\bibitem[{{Fall} \& {Zhang}(2001)}]{fz01}
{Fall}, S.~M. \& {Zhang}, Q. 2001, \apj, 561, 751

\bibitem[{{Figer} {et~al.}(2006){Figer}, {MacKenty}, {Robberto}, {Smith},
  {Najarro}, {Kudritzki}, \& {Herrero}}]{rsg1}
{Figer}, D.~F., {MacKenty}, J.~W., {Robberto}, M., {et~al.} 2006, \apj, 643,
  1166

\bibitem[{{Gallagher} \& {Smith}(1999)}]{gs99}
{Gallagher}, III, J.~S. \& {Smith}, L.~J. 1999, \mnras, 304, 540

\bibitem[{{Goodwin} \& {Bastian}(2006)}]{gb06}
{Goodwin}, S.~P. \& {Bastian}, N. 2006, \mnras, 373, 752

\bibitem[{{Ho} \& {Filippenko}(1996)}]{hp96}
{Ho}, L.~C. \& {Filippenko}, A.~V. 1996, \apj, 472, 600

\bibitem[{{Larsen} {et~al.}(2004){Larsen}, {Brodie}, \& {Hunter}}]{larsen04}
{Larsen}, S.~S., {Brodie}, J.~P., \& {Hunter}, D.~A. 2004, \aj, 128, 2295

\bibitem[{{Lee} \& {Skillman}(2004)}]{ls04}
{Lee}, H. \& {Skillman}, E.~D. 2004, \apj, 614, 698

\bibitem[{{Leitherer} {et~al.}(1999){Leitherer}, {Schaerer}, {Goldader},
  {Gonz{\'a}lez Delgado}, {Robert}, {Kune}, {de Mello}, {Devost}, \&
  {Heckman}}]{leitherer99}
{Leitherer}, C., {Schaerer}, D., {Goldader}, J.~D., {et~al.} 1999, \apjs, 123,
  3

\bibitem[{{Levesque} {et~al.}(2006){Levesque}, {Massey}, {Olsen}, {Plez},
  {Meynet}, \& {Maeder}}]{levesque06}
{Levesque}, E.~M., {Massey}, P., {Olsen}, K.~A.~G., {et~al.} 2006, \apj, 645,
  1102

\bibitem[{{Ma{\'{\i}}z-Apell{\'a}niz}(2001)}]{ma01}
{Ma{\'{\i}}z-Apell{\'a}niz}, J. 2001, \apj, 563, 151

\bibitem[{{Marlowe} {et~al.}(1995){Marlowe}, {Heckman}, {Wyse}, \&
  {Schommer}}]{marlowe95}
{Marlowe}, A.~T., {Heckman}, T.~M., {Wyse}, R.~F.~G., \& {Schommer}, R. 1995,
  \apj, 438, 563

\bibitem[{{Massey} \& {Olsen}(2003)}]{mo03}
{Massey}, P. \& {Olsen}, K.~A.~G. 2003, \aj, 126, 2867

\bibitem[{{McCrady} {et~al.}(2003){McCrady}, {Gilbert}, \&
  {Graham}}]{mccrady03}
{McCrady}, N., {Gilbert}, A.~M., \& {Graham}, J.~R. 2003, \apj, 596, 240

\bibitem[{{Melnick} {et~al.}(1985{\natexlab{a}}){Melnick}, {Moles}, \&
  {Terlevich}}]{melnick85a}
{Melnick}, J., {Moles}, M., \& {Terlevich}, R. 1985{\natexlab{a}}, \aap, 149,
  L24

\bibitem[{{Melnick} {et~al.}(1985{\natexlab{b}}){Melnick}, {Terlevich}, \&
  {Moles}}]{melnick85b}
{Melnick}, J., {Terlevich}, R., \& {Moles}, M. 1985{\natexlab{b}}, \rmxaa, 11,
  91

\bibitem[{{Mengel} {et~al.}(2002){Mengel}, {Lehnert}, {Thatte}, \&
  {Genzel}}]{mengel02}
{Mengel}, S., {Lehnert}, M.~D., {Thatte}, N., \& {Genzel}, R. 2002, \aap, 383,
  137

\bibitem[{{Mengel} {et~al.}(2001){Mengel}, {Lehnert}, {Thatte},
  {Tacconi-Garman}, \& {Genzel}}]{mengel01}
{Mengel}, S., {Lehnert}, M.~D., {Thatte}, N., {Tacconi-Garman}, L.~E., \&
  {Genzel}, R. 2001, \apj, 550, 280

\bibitem[{{Meurer} {et~al.}(1992){Meurer}, {Freeman}, {Dopita}, \&
  {Cacciari}}]{meurer92}
{Meurer}, G.~R., {Freeman}, K.~C., {Dopita}, M.~A., \& {Cacciari}, C. 1992,
  \aj, 103, 60

\bibitem[{{Meurer} {et~al.}(1995){Meurer}, {Heckman}, {Leitherer}, {Kinney},
  {Robert}, \& {Garnett}}]{meurer95}
{Meurer}, G.~R., {Heckman}, T.~M., {Leitherer}, C., {et~al.} 1995, \aj, 110,
  2665

\bibitem[{{O'Connell} {et~al.}(1994){O'Connell}, {Gallagher}, \&
  {Hunter}}]{oc94}
{O'Connell}, R.~W., {Gallagher}, III, J.~S., \& {Hunter}, D.~A. 1994, \apj,
  433, 65

\bibitem[{{O'Connell} {et~al.}(1995){O'Connell}, {Gallagher}, {Hunter}, \&
  {Colley}}]{oc95}
{O'Connell}, R.~W., {Gallagher}, III, J.~S., {Hunter}, D.~A., \& {Colley},
  W.~N. 1995, \apjl, 446, L1

\bibitem[{{Origlia} {et~al.}(1993){Origlia}, {Moorwood}, \& {Oliva}}]{ori93}
{Origlia}, L., {Moorwood}, A.~F.~M., \& {Oliva}, E. 1993, \aap, 280, 536

\bibitem[{{Smith} \& {Gallagher}(2001)}]{sg01}
{Smith}, L.~J. \& {Gallagher}, J.~S. 2001, \mnras, 326, 1027

\bibitem[{{Sternberg}(1998)}]{sternberg98}
{Sternberg}, A. 1998, \apj, 506, 721

\bibitem[{{Storchi-Bergmann} {et~al.}(1994){Storchi-Bergmann}, {Calzetti}, \&
  {Kinney}}]{storchi94}
{Storchi-Bergmann}, T., {Calzetti}, D., \& {Kinney}, A.~L. 1994, \apj, 429, 572

\bibitem[{{Tosi} {et~al.}(2001){Tosi}, {Sabbi}, {Bellazzini}, {Aloisi},
  {Greggio}, {Leitherer}, \& {Montegriffo}}]{tosi01}
{Tosi}, M., {Sabbi}, E., {Bellazzini}, M., {et~al.} 2001, \aj, 122, 1271

\bibitem[{{V{\'a}zquez} {et~al.}(2004){V{\'a}zquez}, {Leitherer}, {Heckman},
  {Lennon}, {de Mello}, {Meurer}, \& {Martin}}]{vazquez04}
{V{\'a}zquez}, G.~A., {Leitherer}, C., {Heckman}, T.~M., {et~al.} 2004, \apj,
  600, 162

\bibitem[{{Wallace} \& {Hinkle}(1997)}]{wh97}
{Wallace}, L. \& {Hinkle}, K. 1997, \apjs, 111, 445

\bibitem[{{Whitmore} \& {Schweizer}(1995)}]{whitmore95}
{Whitmore}, B.~C. \& {Schweizer}, F. 1995, \aj, 109, 960

\end{thebibliography}



\end{document}